# THE SEARCH FOR DELIBERATE INTERSTELLAR SETI SIGNALS MAY BE FUTILE

**JOHN GERTZ** Zorro Productions, 2249 Fifth St., Berkeley, California CA 94710 USA

**Email** jgertz@zorro.com

For more than 60 years, the predominant SETI search paradigm has entailed the observation of stars in an effort to detect alien electromagnetic signals that deliberately target Earth. However, this strategy is fraught with challenges when examined from ET's perspective. Astronomical, physiological, psychological, and intellectual problems are enumerated. Consequently, ET is likely to attempt a different strategy in order to best establish communications. It will send physical AI robotic probes that would be linked together by a vast interstellar network of communications nodes. This strategy would solve most or all problems associated with interstellar signaling.

## 1 INTRODUCTION

Despite much speculation about other possible technosignatures, most actual SETI observation time has entailed the pointing of optical or radio telescopes at individual stars, or multiple stars within a single field of view, in the hopes of detecting electromagnetic (EM) signals that are intentionally being transmitted toward Earth or toward a target in Earth's foreground or background and intercepted by Earth based telescopes. EM transmissions could be without content, perhaps signifying an invitation to a dialogue, or alternatively, they might be information rich. Although much has been written about ET's possible motivations in transmitting signals, little sustained thought has been devoted to looking at the challenges ET would face in attempting to communicate with an unknown civilization such as our own by means of EM signals transmitted from its home solar system. It is the position of this paper that the astronomical, physiological, psychological, and intellectual challenges would be so daunting, and the act of revealing one's location so potentially dangerous, that aliens would turn instead to physical probes as their only viable option for the initiation of interstellar two-way communication.

## 2 INTERSTELLAR EM SIGNALS

From Earth's point of view, we desire that ET persistently transmit the entire body of its knowledge at a flux that is bright enough for us to not only detect its carrier wave, but also to discern its imbedded message. We also hope that ET will provide us with an understandable set of decoding instructions. However, when considering the problem from ET's point of view, our hopes may dissolve into fatally flawed wishful thinking. The problems are so legion that ET might not attempt the enterprise.

### 2.1 Overlapping in Deep Time

The first planets with sufficient metallicity to become Earth analogs may have been formed as early as 12 billion years ago. Approximately 95% of all currently extant stars are older than the Sun. Consequently, the first alien civilizations capable of communicating with Earth may have formed long before the birth of our Solar System. It is safe to assume that aliens know at least the age of the Earth. We know that the first technologically capable species evolved on Earth 4.5 billion years after its formation. We do not know how far this is from the average time it takes for a technological species to evolve in accordance with models that may be available to ET. If the average is, for example, 4.5 billion years with a 2 SD interval bar of one billion years, then ET might have begun transmissions to Earth, according to our wishful thinking, about one billion years ago. That is a very long time to expect ET to persistently target Earth in the anticipation that a technologically competent species, such as homo sapiens, might eventually emerge. ET itself must persist during this long duration, with an unwavering mission to make contact with Earth, despite its own biological or post-biological evolution (not to mention its possible extinction in the interval), changes to its environment, its religion, its politics, or whatever else might drive it.

### 2.2 A Dedicated Transmitter in Current Time

Assuming that ET and homo sapiens coexist as transmitting and receiving civilizations in deep time, either ET will require a transmitter dedicated persistently to Earth or we will require a telescope dedicated to only ET's star system. Since the latter is certainly not the case, the onus is upon ET. The problem is somewhat ameliorated if one or the other telescope targets multiple stars simultaneously through beam forming, however those stars must all be within the telescopes primary field-of-view. If either the transmitting or receiving civilizations are primarily interested in the nearest stars then those will be randomly distributed across the sky and not eligible for simultaneous observation with multiple beams. All-sky-all-the-time systems would likely be too insensitive to do more than detect a bright carrier signal, but would likely be inadequate for de-coding imbedded messages. SETI scientists sequentially target long lists of stars, with a typical dwell time of about ten minutes and a duty cycle of many years. If ET transmits with the same cadence, the chances of ET's transmission and Earth's observation aligning in current time is vanishingly small.

### 1.3 Planning for Earth's Response

Whether or not ET transmits to Earth persistently, it must dedicate a receiver to Earth's response. If, for example, ET lies at the distance of 100 light years, every time it transmits to Earth it must have a receiver ready 200 years later. But how long should that receiver remain focused on Earth? Surely ET will understand that it will take time for the recipient civilization to decode the message and decide whether and what to respond. Should ET wait a month, a year, more? In the event that ET employs a long target list, then it will have to have many dedicated receivers at the ready, one for each star system to which it transmits. If ET's strategy is the same as Earth's, namely, to sequentially target many stars, each for ten minutes, then it must also dedicate some period of time (as adjusted by distance in light years) for a response from each target. The number of receivers that would be necessary for this job would dwarf the number of necessary transmitters.

### 1.4 Timeliness of the Message

Everything about which we would wish to communicate with ET changes as a function of our historical moment. Perhaps it is the same with ET. Since I cannot use knowledge of our future, I will use the past to illustrate the point. If we had established communication with an exo-planet in the Kepler field (~1,500 light years from Earth) some 3,000 years ago, we might just now be receiving ET's response. We might have sent our most advanced literature, Gilgamesh; our most advanced math, addition and subtraction without any reference to zero, which had not yet been invented; advice on how to build the highest ziggurat; and instructions on how to worship the sun god, Ra. ET might not send actually useful math or science in return, since it might reason that we are not yet able to understand relativity and quantum physics, much less anything more advanced than that. We would have only signalled our pathetic ignorance and unworthiness as a communicating partner; 3,000 years hence we might be deeply embarrassed by what we chose to transmit at this time.

### 1.5 Power Games

The standard SETI observing protocol makes the further tacit assumption that ET is transmitting not only persistently in both deep time and current time, but doing so with a very powerful transmitter at great and sustained costs in energy and materials. But why ever would it? ET may make the opposite assumption, that the onus is upon the recipient to build a very large receiver. That should only be fair and right. After all, ET is giving the recipient the benefit of its transmission, which may contain the entirety of its knowledge and wisdom. It is therefore incumbent upon the recipient to build a gargantuan telescope to properly receive it [1]. This may be all the more true because ET will realize from the outset that it may have to transmit over potentially a billion years or more as it waits for a technologically competent species to evolve on the target planet, while that species need only scan the skies for a relatively brief period of its history, perhaps a hundred years or less, before it detects ET's persistent signal. Of perhaps even more importance to the aliens is that while the act of transmitting requires enormous energy, the act of receiving requires almost none. By analogy, by means of its vast energy emissions a star is observed by a telescope, which in turn requires little more than the energy to run a camera, some orientation gears, and some back-end computing. The transmitter/receiver relationship is linear. The larger or energy intensive the transmitter, the smaller the receiver required to detect its transmission, and vice versa. ET can save enormously on energy by building only a humble transmitter that would not be detectable by Earth's current generation of telescopes. Perhaps ET does transmit all of its knowledge and wisdom in the hopes of receiving ours in return. Nevertheless, good trade must be fair trade, and any sense of fairness dictates that the burden should be on us to build gargantuan telescopes rather than on ET to build gargantuan transmitters. If ET is indeed transmitting at low or even modest energies, then our current generation of telescopes are simply inadequate to achieve a detection.

### 1.6 This Is Dangerous; Don't Try It at Home

ET may know for a fact, or have reason to suspect, that the galaxy contains some aggressive and dangerous civilizations or AI beings. In such event, interstellar EM transmissions might alert hostile forces to ET's existence, inviting a disastrous response. This alone may dissuade it from attempting to use EM transmissions as a means of intergalactic communication.

### 1.7 Technical Impediments

Interstellar EM transmissions are beset by daunting technical limitations that make intergalactic communications difficult, if not impossible. For example, lasers are very efficient potential carriers of information. By means of photomultipliers attached to large terrestrial telescopes, pulsed bit rates of a billion or more per second might be accurately recorded. However, because of intervening dust, lasers damp down significantly with distance, such that in most lines of sight they would dim to invisibility within 1,000 light years.

Interstellar radio transmissions are subject to dispersion and scattering due to inhomogeneities in electron densities and turbulence in the interstellar medium (IM), resulting in a fading in and out of the message and distortions of wave forms and pulse timing. As a consequence, messages may be garbled beyond the ability of the receiving civilization to accurately record and decode [2, 3].

## 3 COMPATIBILITY

The dominant SETI research protocol demands that ET expend an immense amount of energy over deep time to persistently transmit to Earth in the absence of any specific knowledge of Earth's emergent technologically competent species. However, ET would have to make unwarranted assumptions of mutual compatibility that on consideration would be so daunting as to dissuade it from any such effort in the first place. The possibilities for miscommunication would be vastly larger than that for mutual comprehensibility.

### 3.1 Incompatible Sensory Systems

Homo sapiens rely on a discrete set of sensory systems to understand the world around it (sight, sound, touch, and so forth). Aliens may rely on different sensory systems or on homologous but differently weighted sensory systems. They may see infrared or ultraviolet or may be color blind or have no sight at all. They may understand their world through sensations of electrical or magnetic fields. Dogs and humans each hear and smell, but each relies on its senses in different ways, humans to discern music and dogs to smell the urine of other dogs.

### 3.2 Incompatible Modes of Communication

We communicate by means of facial expressions, speech, pictures and written language. Aliens might communicate with bee-like tummy waggles, dolphin-like squeaks and whistles, ant-like pheromones, bat-like echolocations, cuttlefish-like color transitions, or in ways we cannot even imagine. They may lack ears (or their homologues) or lack any ability to understand that subtle air oscillations made by a food intake organ can also be coopted for the purpose of communication. Of course, there is written language, but that too is fraught with problems. We still cannot decode certain ancient written languages such as Indus and Etruscan, primarily because we lack a Rosetta Stone, some manner in which one language is related to another through common referents. Even if ET has a written language, if the referents are to changes in a magnetic field that convey meaning to them just as pictures convey meaning to us, successful communication between us may be nearly impossible.

### 3.3 Intellectual Incompatibility

Even if the transmitting and receiving civilizations share comparable sensory systems and modes of communication, there may be an unbridgeable intellectual gap between the two. Homo sapiens and homo habilis are examples of species that share almost everything except brain size. Consequently, it would be pointless for the former to attempt to share Shakespeare or Einstein with the latter. Having no a priori understanding of our intellectual capacities, ET would hardly know how to calibrate its message.

### 3.4 The Use of Math and Science as Stepping Stones Toward Comprehensibility

Perhaps mathematics might form a natural language for alien communication. We do not know whether mathematics is inherent in the cosmos, or if it is instead a human construct that we have ourselves created in order to understand the world. Most mathematicians have the sense that they are discovering math rather than inventing it, and that math is the driver by which the universe was created and works. In that case, perhaps ET's math is similar to, though presumably more advanced than, our own.

Perhaps ET will commence its communications with rudimentary mathematical concepts, such as 1+2=3, and build up from there. From simple building blocks it might soon arrive at formulas such as $E=mc^2$, at least mathematically. But to understand what this equation is saying, one needs to first have conveyed the concepts of energy, mass, and the speed of light, or at least two of the three concepts. Brian McConnell suggests that physical concepts can be taught by sending computer codes that illustrate physical concepts. For example, an algorithm can be devised to simulate objects behaving in a gravitational field [4].

Some physical concepts can be conveyed through ratios. For example, protons have 1,836 times the mass of electrons, regardless of what idiosyncratic measurements are employed. Therefore, the number 1,836 can be used to convey that the objects referred to are these particles [5]. Once math has been fully explicated, along with as much physics as might be conveyed thereby, a bridge to chemistry might be devised. Schemes have been suggested to communicate fundamental understanding of chemistry, starting with the periodic table [6]. From chemistry, some basic biology, such as DNA or RNA codes, or the chemiosmotic process might be possible to transmit [7]. Only rudimentary work has been done in trying to devise such stepwise building of languages [8,9]. Methods for conveying emotional and cultural constructs, such as love, beauty, democracy, racism, or fun, using a language constructed from math, has been barely explored, and may in fact not be possible. Perhaps some progress might be made by teaching first the mathematical symbols = and ≠, as in 1 + 1 = 2 and 1 + 1 ≠ 3, illustrating mathematically the concepts of correct and incorrect. This might then be expanded to apply to the moral concepts of right versus wrong. This might be done by showing a picture of one person greeting another person with the = sign, and then another picture of a person striking another person with the ≠ sign [10]. But such a system is fairly rudimentary and still does not take us all the way to concepts of divinity, much less a respectable translation into lingua cosmica of the Bible, the Upanishads, or the I Ching. Aliens may not wish to expend all of the deep time and immense energy to communicate with other civilizations simply to discuss their math and science, especially if they suspect that theirs is far in advance of any species on a planet that is a mere 4.5 billion years old. The thing that might interest them most are just those subjects that are hardest, and perhaps even impossible to convey.

### 3.5 Incompatible Subjects of Concern

Our subjects of concern are tantamount to our culture and include literature, music, art, history, politics, philosophy, religion, and so forth. Alien interests may differ so greatly from ours that we might only be able to call them "concerns," or "subjects of focus," rather than "culture." My dog fixates on the smells of each bush we walk by. I have no idea what world of smells she is entering. If my dog could speak, perhaps she would describe and have names for 50,000 different bush smells, each conveying much meaning to her, but of which I am completely insensible. Even if she could poetically describe to me the glorious pleasure she derives from the subtle bouquets of splotched dog urine, I doubt I could ever learn to appreciate them. On the other hand, when I listen to music, my dog seems to be entirely indifferent and unable to differentiate between Bach and the Backstreet Boys, or Haydn and hip hop, or for that matter to care in the slightest. This is not a difference between us in intelligence; it is a fundamental difference in concerns. Dogs and homo sapiens are both mammals living on the same planet, who have co-evolved for at least 10,000 years and probably much longer. How much vaster might be the gulf between ourselves and aliens? Culturally, it may be all in the sensory receptors of the beholder. They may have no appreciation whatsoever for Shakespeare, and we may have none for the subtlety of their latest waggle dancing or pheromone-based art.

### 3.6 Misunderstandings Might Outweigh Understandings

Even in the very unlikely event that aliens and homo sapiens share compatible sensory systems, modes of communication, intellectual capacities, math, science, and subjects of concern, any communication across interstellar distances would be ripe for misunderstanding, with possibly catastrophic results. For example, in 1977, Frank Drake, Carl Sagan and Anne Druyan affixed plaques to the sides of Pioneer 10 and 11, intended for whatever aliens might at some point intercept the spacecraft in the depths of interstellar space.

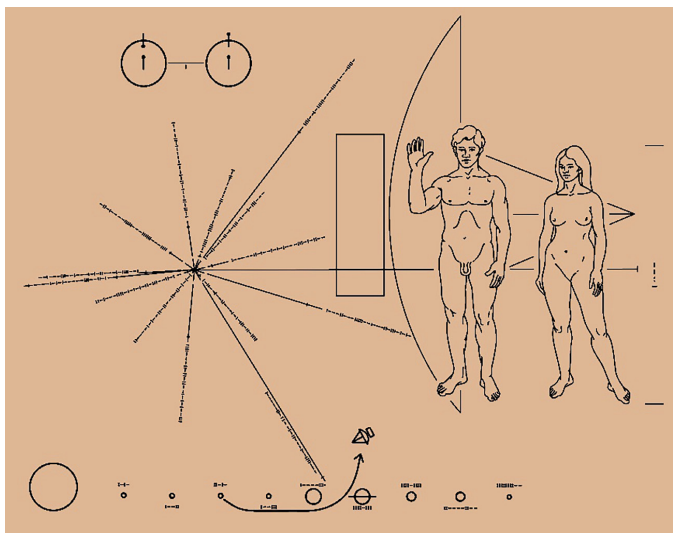

**Fig.1** The plaque affixced to Pioneers 10 and 11, devised in 1977 by Frank Drake, Carl Sagan and Anne Druyan.

The nakedness of the two figures could be understood to mean that the two pictured beings have no claws, fangs, or horns, and are therefore extra dangerous because they rely on nuclear explosions (seen at left) for offense, and shields (behind the beings) for defense. An upraised appendage might signify to aliens a grave insult, much like a raised middle finger. The two figures might bear a vague resemblance to a species of vermin on the alien's home planet. The row of orbs below might signify an interstellar catapult by the means of which Earthlings intend to launch their bombs against ET. In short, the plaque might be taken as a declaration of war rather than of peace (one can refer to Wikipedia or elsewhere on the Internet to learn the plaque's actually intended meaning). Any communication undertaken before one civilization has thoroughly studied the other may be counterproductive and fraught with opportunities to fatally misunderstand one another.

### 4   PROBES LINKED BY NODES ARE ET'S BETTER, AND POSSIBLY ONLY, OPTION

Building on the work of others, the author has hypothesized that ET's better strategy for communicating across interstellar distances would be to send physical robotic probes to distal star systems [11–21]:

- An alien probe(s) with AI capabilities placed within our own Solar System might take whatever time it requires to surveil Earth's omnidirectional EM leakage. It might learn our language from Sesame Street, our math from Khan Academy, and everything else from YouTube. It would commence communication only once it has thoroughly decoded us.
- Because the local probe(s) would have only a single transmission target, Earth, once it does commence transmissions, they will be persistent.
- Because the local probe(s) will be situated at a distance of light seconds or light minutes rather than at a distance of hundreds or thousands of light years, even a small on-board transmitter will likely transmit at a received flux that would be orders of magnitude brighter than putative interstellar EM transmissions.
- Because it is local, it would not be subject to such complicating effects as dimming (lasers) or dispersion (radio).
- Due to its proximity, a local probe would enter into a dialogue in near real time, solving for the problem of timeliness and relevance of the response.
- Because the probe has surveilled Earth's EM leakage for decades at least, when it does deign to transmit to Earth, it might do so in English or some other terrestrial language that it had learned, immeasurably increasing the likelihood that we will understand the content of the communication.
- There may be a great Internet of communication probes (i.e., nodes) connecting all member civilizations, facilitating the dissemination of information about Earth and homo sapiens throughout the galaxy.
- The probe will be able to communicate its findings back to its progenitor civilization and to all other member civilizations, translating homo sapiens into terms that might be understandable to aliens.
- A local probe might have been launched prior to the demise of its progenitor civilization, solving for the coexistence problem in deep time, or Drake's L.
- A local probe need not reveal the coordinates of its progenitor civilization(s), thus eliminating the inherent danger posed to aliens were they to seek communication partners by means of interstellar EM transmissions.
- A local probe can do actual science and research even before the emergence of a technologically competent civilization. Interstellar EM transmissions return no information in the absence of contact with another civilization.
- When encountering a technologically competent civilization, a local probe can assess any potential danger to the galaxy, that is, it can yield situational awareness to the progenitor civilization(s).
- The local probe may have at its disposal aggressive means to deal with any newly emergent technologically competent civilization that is deemed to be dangerous to the progenitor civilization or the galactic club of peaceful civilizations. In fact, defense may be a prime imperative for the placement of robotic probes within biogenic star systems where they can observe a species closely as they emerge into technological competence. It must surely be easier to strangle the baby in its crib than to wait for a new civilization to develop interstellar offensive capabilities. Unbeknownst to us, we may be undergoing a judgment by AI robotic probes that are fully capable of destroying us if their algorithms so deem it to be necessary.

### 5   INTERSTELLAR SETI MAY NOT BE COMPLETELY FUTILE

The predominant SETI paradigm, observing for intentionally transmitted interstellar EM transmissions, may be so fraught that aliens would never attempt it. They would favor a much safer and much more efficient strategy, employing physical probes linked together by communication nodes. However, not all interstellar SETI is futile. There are three major exceptions.

#### 5.1   The "I Love Lucy" Exception

Had alien civilizations residing within about 70 light years from Earth pointed sufficiently large radio telescopes toward Earth, they might have detected Earth's earliest persistent terrestrial television signals. In such a case, those residing within 35 light years would have had time to respond. Approximately 750 stars lie within this radius, a tiny number in a galaxy of over 200 billion stars. Nonetheless, this small set of stars form a special case, worthy of routine and regular observation. Moreover, the radius might be increased a bit if one considers that aliens might have detected other techosignatures even earlier than circa 1950, such as city lights, a rise in carbon dioxide at the outset of the Industrial Age, or the residues of atmospheric nuclear explosions from the mid-1940s.

## 5.2 The K-III Exception

Aliens that are capable of harnessing substantial portions of their galaxy's entire energetic output might be able to transmit signals intergalactically (so-called K-III civilizations). It would presumably be more efficient to illuminate an entire distal galaxy at once with beamed transmissions than to send a fleet of physical probes. Moreover, safety concerns would diminish because would-be hostile actors would be at a great disadvantage in attempting to aggress across vast intergalactic distances, and K-III civilizations might be quite secure in their very advanced capabilities to defend themselves [22].

## 5.3 The Search for Technosignatures

Although not intentionally transmitting signals from their home solar system, aliens may nevertheless be detectable through such techosignatures as their city lights, the infrared waste heat of their artificial structures (so-called Dyson spheres), artificial molecules in their atmospheres, nuclear wastes deposited in their star, the odd shapes of their orbital megastructures as they transit their stars as seen from Earth, left physical artifacts within our own Solar System, and so forth.

## 6 CONCLUSIONS

We currently perceive a Great Silence, the apparent absence of alien interstellar EM transmissions deliberately targeting Earth. This may simply reflect the fact that aliens have determined that interstellar signaling is not merely a sub-optimal strategy when compared to sending physical probes for the purpose of surveilling and making contact with other civilizations, but that it may be dangerous and completely unworkable due to the astronomical, physiological, psychological, and intellectual problems enumerated in this paper. ET may then have determined that its best and, in fact, only viable strategy for communication would be through a vast galactic network of communication probes and nodes that might disseminate its findings among member civilizations. Probes and nodes might shunt partially analyzed data in the direction of one or more civilizations with the closest match in compatibility to the target civilization for deeper analysis as well as, perhaps, for instructions on how to proceed: (a) initiate contact; (b) remain passive while collecting more data, or (c) destroy forthwith a civilization deemed to be dangerous.

## Acknowledgement

The author gratefully acknowledges the kind assistance of Geoff Marcy.